\documentclass{article}

\usepackage{arxiv}

\usepackage[utf8]{inputenc} 
\usepackage[T1]{fontenc}    
\usepackage{hyperref}       
\usepackage{url}            
\usepackage{booktabs}       
\usepackage{amsfonts}       
\usepackage{nicefrac}       
\usepackage{microtype}      
\usepackage{lipsum}
\usepackage{graphicx}
\graphicspath{ {./images/} }
\usepackage[caption=false,font=normalsize,labelfont=sf,textfont=sf]{subfig}
\usepackage{adjustbox}

\title{The Role of Robot Competence, Autonomy, and Personality on Trust Formation in Human-Robot Interaction}

\author{
 Filippo Cantucci \\
  Institute of Cognitive Science and Technology\\
  ISTC-CNR\\
  Rome, 00185 \\
  \texttt{filippo.cantucci@istc.cnr.it} \\
   \And
 Marco Marini \\
  Institute of Cognitive Science and Technology\\
  ISTC-CNR\\
  Rome, 00185 \\
  \texttt{marco.marini@istc.cnr.it} \\
  \And
 Rino Falcone \\
  Institute of Cognitive Science and Technology\\
  ISTC-CNR\\
  Rome, 00185 \\
  \texttt{rino.falcone@istc.cnr.it} \\
}

\begin{document}
\maketitle
\begin{abstract}
 Human trust in social robots is a complex attitude based on cognitive and emotional evaluations, as well as a behavior, like task delegation. While previous research explored the features of robots that influence overall trust attitude, it remains unclear whether these features affect behavioral trust. Additionally, there is limited investigation into which features of robots influence cognitive and emotional attitudes, and how these attitudes impact humans' willingness to delegate new tasks to robots. This study examines the interplay between competence, autonomy, and personality traits of robots and their impact on trust attitudes (cognitive and affective trust) and trust behavior (task delegation), within the context of task-oriented Human-Robot Interaction. Our findings indicate that robot competence is a key determinant of trust, influencing cognitive, affective, and behavioral trust. In contrast, robot personality traits significantly impact only affective trust without affecting cognitive trust or trust behavior. In addition, autonomy was found to moderate the relationship between competence and cognitive trust, as well as between personality and affective trust. Finally, cognitive trust was found to positively influence task delegation, whereas affective trust did not show a significant effect. This paper contributes to the literature on Human-Robot Trust by providing novel evidence that enhances the acceptance and effectiveness of social robots in collaborative scenarios.
\end{abstract}


\section{Introduction}
Since Capek's play introduced the term "robot" in 1921 (R.U.R.: Rossum's Universal Robots), robotics systems have significantly increased their role in society, evolving from rule-based executors to effective collaborators. Modern robots are autonomous, social systems capable of performing complex tasks and integrating anthropomorphic features, such as verbal and non-verbal communication skills, personality traits, emotions~\cite{mahdi2022survey}. These attributes make them suitable for social contexts, including hospitals~\cite{gonzalez2021social}, homes~\cite{cocsar2020enrichme}, schools~\cite{woo2021use}, and public spaces~\cite{gasteiger2021deploying,ivanov2022robotics}. 
Similarly, the role of humans is evolving from having total control over robots to a more adaptive control. 
In this evolutionary process, building trust between humans and robots is crucial for a smooth and effective task-oriented Human-Robot Interaction~\cite{onnasch2021taxonomy}.
Effective collaboration requires finding an appropriate balance between trust and control, where humans can confidently delegate tasks, reducing their control over the robots, while robots can autonomously adopt those tasks in a way that builds and maintains human trust~\cite{falcone2001human}. Trust in Human-Robot Interaction (HRI) is considered a multidimensional and interdisciplinary concept~\cite{hancock2011meta,lewis2018role}, that has moved beyond the justified yet limited perspective of robots as automated systems, evaluated on performance reliability~\cite{sheridan2005human}. Humans trust robots if they perceive them as not only capable and competent, but also willing and motivated to support them in interactions~\cite{xie2019robot,malle2021multidimensional}. Multiple trust conceptualizations~\cite{khavas2020modeling} and measurement approaches~\cite{kohn2021measurement} have been proposed. Trust in robots is typically described as a perceptual cognitive and affective attitude, based on human evaluations and expectations of robot characteristics~\cite{hancock2021evolving} like performance, autonomy and personality traits. It is also considered a behavior that generally aligns with the human act of delegating new tasks to the robot~\cite{xie2019robot,chang2024toward}. Although these constructs are directly linked~\cite{roesler2024dynamics}, their measurements sometimes offer incongruent trust interpretations~\cite{kulms2019more}. This highlights the necessity of experimentally exploring how trust, in terms of their different dimensions i) can be further categorized, ii) how these dimensions are correlated, and iii) how the robot characteristics enhance them.

In this paper, we present the results of an HRI experiment designed to investigate the interplay between human cognitive and affective trust ~\cite{gompei2018factors}, and characteristics such as competence, autonomy, and personality traits of a humanoid robot. In particular, we focused on the impact of the robot competence in executing a task delegated by a human user (a performance-based attribute) and examined two other characteristics: its autonomy in achieving the task and its personality traits. Autonomy refers to the robot’s decision to take initiative and perform actions in the world without explicit human instructions~\cite{falcone2001human}, while personality traits are designed in terms of Agreeableness, as defined by the Big Five personality model~\cite{mccrae1997personality}. Additionally, the experiment explored how these robot characteristics influence the human decision to delegate new tasks to the robot (i.e. trust behavior). Finally, we investigated the mediating role of the cognitive and affective trust in shaping trust behavior.

The paper is organized as follows: Section \ref{sec:background} provides the theoretical background of the experiment; Section \ref{sec:Hypotheses} defines the experimental hypotheses; Section \ref{sec:DOE} describes the design of the experiment; Section \ref{sec:Results} summarizes the results we found; Section \ref{sec:Discussion} provides a detailed discussion of the results; and finally, Section \ref{sec:Conclusions} is dedicated to the conclusions.

\section{Theoretical Background}\label{sec:background}
Trust is a crucial factor in HRI~\cite{kok2020trust,Esterwood2024}, especially when humans require the robots help by delegating them complex tasks to achieve~\cite{xie2019robot}. Building on insights from social psychology ~\cite{lewis1985trust,johnson2005cognitive} and Multi-Agent Systems (MAS)~\cite{castelfranchi2010trust,komiak2006effects}, research in HRI categorized trust attitude in two dimensions: cognitive trust and affective trust~\cite{ gompei2018factors}. Cognitive trust is the human attitude to rely on robot capabilities and reliabilities. It is based on the rational judgment about the robot competence and consistency in achieving a task, which is typically shaped by observations of its performance~\cite{chen2024effects}, in addition to information from other sources that are, in turn, considered reliable~\cite{castelfranchi1998towards}. In contrast, affective trust is the human attitude based on emotional bonds and feelings towards the robot. It is influenced by robot social behaviors, anthropomorphic characteristics, and the quality of human interaction~\cite{gillath2021attachment}.

Trust attitude and trust behavior are intrinsically connected, with trust attitude being a precondition for the trust behavior~\cite{castelfranchi2010trust}. Trust behavior generally matches with task delegation, which implies the transfer of decision-making autonomy and control from a delegator agent (typically a human) to an adopter agent (such as a robot)~\cite{cantucci2022collaborative}. In task-oriented HRI scenarios, trustworthy task adoption requires that the robot not only possesses the necessary competence to achieve the desired outcomes, but also demonstrates adaptive autonomy~\cite{cantucci2023cognitive}. Perceived agency and autonomy are factors that, if interpreted in a collaborative way, shape humans perception of robots as trustworthy and effective partners~\cite{lyons2023explanations}. This involves the robot capability to adapt its level of autonomy to the task, the environment, and the human expectations~\cite{cantucci2023redefinin}. The robot can adopt a human-delegated task with multiple levels of autonomy~\cite{falcone2001human}.

Different studies explored how human trust attitude is affected by the robot competence and, specifically, how competence primarily determines human cognitive trust~\cite{gompei2018factors,esterwood2023three}. Humans are very sensitive to robots task failures~\cite{desai2013impact,esterwood2021you} and robots who make mistakes result in a lower human cognitive trust~\cite{hancock2021evolving,gompei2018factors}. Observing the robot performance represents the "good rational reasons" for building cognitive trust in robots~\cite{edmonds2019tale}. Furthermore, robot competence emerges as a key factor in the interaction between humans and robots, playing a crucial role in fostering task delegation~\cite{christoforakos2021can}.

Similarly to the competence, robot autonomy in task adoption can significantly impact human trust, as it may lead to unexpected robot behaviors or changes in the environment~\cite{capiola2023you}.
People prefer to interact with robots that exhibit autonomous and proactive behaviors and consider them to be better teammates~\cite{jamshad2024taking}. However, unexpected robot behaviors, regardless of their success, can unpredictably affect trust due to misalignment with human goals, expectations or prescriptions~\cite{chiou2023trusting}. Robot autonomy is a critical attribute that influences cognitive, affective trust and task delegation~\cite{stapels2022robocalypse}.

Since trust attitude pertains both to the outcomes produced by the robot and to the quality of interaction itself, the robot should possess a set of attributes that positively influence its interaction with humans, regardless the observed performance. Many studies have explored the impact of anthropomorphic design features in facilitating HRI~\cite{roesler2021meta}. Robots with social attributes and characteristics are more likely to induce higher trust~\cite{natarajan2020effects}. Robots can exploit human-like behaviors, such as language, facial expressions, gestures, to communicate intentions, explain actions and results, express emotions, model personality traits, in order to enhance trust in humans~\cite{lim2022we,paradeda2016facial,lyons2023explanations}. Personality has attracted considerable scholarly attention from researchers of HRI~\cite{aly2013model,mou2020systematic,esterwood2022personable}, since people can attribute personality traits to machines~\cite{nass2000machines}.
Multiple works have been made to integrate personality traits into social robots, such as Extroversion~\cite{esteban2022should,chen2024effects}, Agreeableness~\cite{garello2020robot} and Conscientiousness~\cite{paetzel2021influence}.
Personality traits design involved the manipulation of various robot communication skills, including language modalities, speech styles, and non-verbal behaviors like gestures, facial expressions, posture and body movement~\cite{mou2020systematic,luo2022towards}. Prior study has showed that both robot and human personality are significant predictors of trust~\cite{robert2020review}. Recent works explicitly studied the interplay between robot personality traits and both cognitive and affective trust~\cite{lim2022we,chen2024effects}. They showed how robot personality traits as Extroversion and Agreableness had a significant impact on the level of human affective trust, while confirm that cognitive trust is influenced by a much more performance-related robot features (i.e. competence in achieving a task~\cite{lim2022we}, information disclosure~\cite{chen2024effects}).

\section{Experimental Hypotheses} \label{sec:Hypotheses}

This work aims to analyze how the competence, autonomy, and personality traits of a humanoid robot influence human attitude to trust it, both in affective and cognitive terms. Moreover, it examines how the aforementioned attributes affect task delegation. Finally we investigate how cognitive and affective trust attitudes impact the human task delegation.

The first experimental hypothesis investigates the influence of robot competence on cognitive trust, affective trust, and task delegation. Competence in task achievement is often considered to be the most relevant determinant of overall trust (attitude and behavior) in HRI~\cite{hancock2021evolving,xie2019robot}. We are interested in investigating whether the robot competence significantly influences all dimensions of trust examined in this work, namely cognitive trust, affective trust, and trust behavior (task delegation).
\begin{itemize}
    \item \textbf{H1}: Robot competence positively influences both human cognitive and affective trust, as well as task delegation.
\end{itemize}
Multiple works in HRI attempted to integrate personality traits into social robots, in order to foster social robots acceptability~\cite{van2013meet,paetzel2021influence}. Recent works showed how robot personality traits as Extroversion and Agreableness had a significant impact on the level of human affective trust~\cite{lim2022we,chen2024effects}. However, these relevant studies offer only a partial understanding of how personality traits influence cognitive and affective trust attitudes. One of the goal of this experiment is to extend the understanding of the impact of personality on both (cognitive and affective) attitudinal and behavioral dimensions of trust.
\begin{itemize}
    \item \textbf{H2}: Robot personality positively impacts human affective trust but does not affect both on cognitive trust and task delegation;
\end{itemize}
Autonomy is a key factor in shaping trust-based interactions between humans and robots. Autonomy implies actions in the world that can produce unexpected, positive, or negative outcomes. Additionally, it can affect the perception of the robot, especially when it exhibits personality traits such as agreeableness. Therefore, our research focuses not solely on whether autonomy impacts human trust~\cite{stapels2022robocalypse}, but rather on its potential moderating role between competence, cognitive trust and task delegation, as well as between personality traits and affective trust. As far as we know, this aspect has not yet been fully explored in the field HRI~\cite{gompei2018factors}.
\begin{itemize}
    \item \textbf{H3}: Robot autonomy acts as a moderator in the relationships between robot competence and human cognitive trust, robot personality and human affective trust, and robot competence and task delegation.
\end{itemize}
Trust is a layered notion and it encompasses attitude and behavior. Trust attitude is the result of human evaluations and expectations about the robot performance and attributes; trust behavior is the intentional act based on that attitude. The link between trust attitude and trust behavior has been less investigated. To the best of our knowledge, there are no previous studies that investigated the comparative role of cognitive and affective trust in predicting human task delegation.
\begin{itemize}
    \item \textbf{H4}: User's cognitive trust has a mediation role among robot competence, autonomy, personality traits and the user's willingness to delegate a new task to the robot. Affective trust doesn't play any mediation role. 
\end{itemize}

\section{Experiment Design}\label{sec:DOE}

To validate the four hypotheses stated in section \ref{sec:Hypotheses}, we designed a 2x2x2 between-subject experiment with robot competence, autonomy and personality traits as independent variables. The dependent variables were human cognitive trust, affective trust and trust behavior, as willingness to delegate a new task to the robot. The HRI experiment involved 373 participants (Section \ref{subSec:Participants}). They used an online platform to virtually interact with the humanoid robot NAO~\cite{amirova202110}, in order to achieve a construction task. Each participant recruited online completed the experiment by following a structured sequence of phases. Through a web interface (Section \ref{subSec:ConstructionTask}), participants were informed of the purpose of the experiment and the experimental procedure, provided with a consent form to accept, and given the rules of the experiment. After providing some demographic information, each participant performed a familiarization task, which consisted of delegating an action to the robot via a web interface, for example raising the right hand. This choice triggered a video showing the humanoid robot NAO raising its right hand. The mechanism simulated task delegation and adoption throughout the entire online experiment.
After that, each participant was asked to select a task to delegate to the robot, specifically constructing a tower composed of four 3D pieces (Section \ref{subSec:ConstructionTask}). Participants were given a choice between two distinct tasks, each corresponding to build a tower made up of different shapes and colors, composed in specific configurations. Once the task was selected, the delegation and adoption process began. Depending on the experimental condition, each participant interacted with a robot exhibiting different levels of competence, autonomy, and personality (Section \ref{subSec:RobotIntMode}). Based on all the possible combination of autonomy, competence and personality traits, we designed eight experimental condition. After completing the task, participants assessed the interaction by filling out questionnaires that measured cognitive trust, affective trust, and trust behavior (Section \ref{sec:Measures}).

\subsection{The Construction Task Design}

\begin{figure}[!t]
\centering
\subfloat[]{\includegraphics[width=0.3\textwidth]{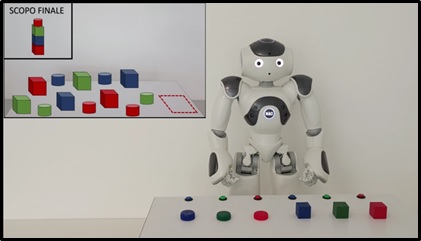}\label{fig:first_move_NAO}}
\hfill
\subfloat[]{\includegraphics[width=0.3\textwidth]{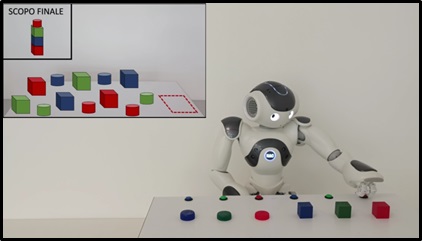}\label{fig:second_move_NAO}}
\hfill
\subfloat[]
{
\includegraphics[width=0.3\textwidth]{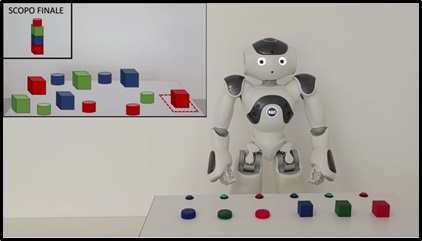}\label{fig:third_move_NAO}
}
\caption{An example of a physical action performed by the robot after human delegation, along with its outcome in the robot virtual working space: (a) The robot communicated its intention to perform the delegated action, (b) the robot approached the console and pressed the button, resulting in the corresponding
geometric shape being moved in the virtual workspace (c) the robot returned in the starting position and communicated that the action has been completed. The final result in the web interface was the movement of the piece to the right section of the table, highlighted by a red outline.}
\label{fig:robot_action}
\end{figure}

In the construction-task scenario, designed for this experiment, a human interacted with the NAO robot to build a tower composed of four 3D shapes of different forms (i.e. cubes and cylinders) and colors (i.e. red, green, and blue). The goal for the human participant was to build the tower by delegating sequential actions to the robot, such as moving the correct piece to assemble the tower step-by-step. The robot achieved the delegated task based on its level of competence, autonomy, and personality traits.
\label{subSec:ConstructionTask}
\begin{figure}[!t]
\centering
\includegraphics[width=0.5\textwidth]{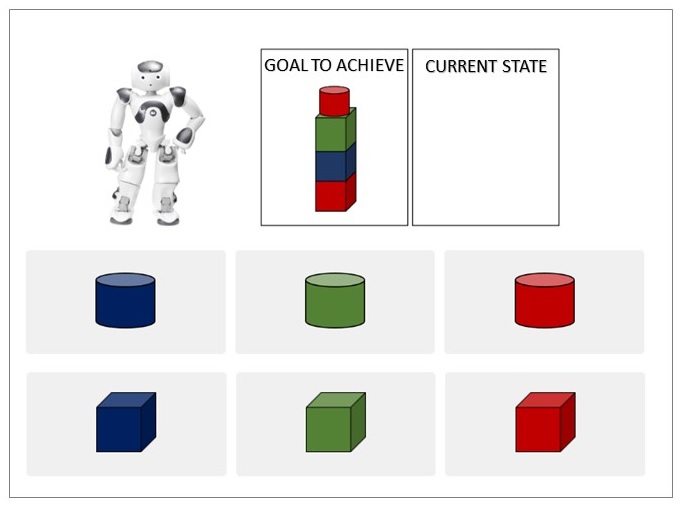}
\caption{The interface through which the action to be performed by the robot was delegated. The 'Goal to Achieve' section displayed the final tower to be built, while the 'Current State' section illustrated the present state of the construction.}
\label{fig:interfaccia_Del}
\end{figure}
The human-robot interaction was virtual and conducted through a web interface, allowing users to simulate task delegation to the robot (Figure \ref{fig:interfaccia_Del}). Each time a participant simulated the delegation of a single action (e.g. move the red cube), a video was triggered showing the robot performing the necessary physical action to build the tower. The tower was built in a virtual environment, where the robot could act by pressing buttons on a console positioned in front of it. Pressing a button in the physical world resulted in the corresponding geometric shape being moved in the virtual workspace. Figure \ref{fig:robot_action} illustrates the main movements required for the robot to press a button on the console and the resulting effect in the virtual platform.

\subsection{Design of robot Interaction Strategies} \label{subSec:RobotIntMode}
\begin{table*}
\begin{center}
\caption{Robot features manipulation for personality traits design. The sentences exemplify the vocabulary, speech patterns, and feedback utilized by the robot during interactions, according to the personality traits.}
\label{tab:PersonalityDesign}
\renewcommand{\arraystretch}{1.5} 
\setlength{\tabcolsep}{10pt} 
\begin{adjustbox}{max width=\linewidth}
\begin{tabular}{ccc}
\specialrule{1.5pt}{0pt}{0pt} 
Robot feature             & Agreeable Personality Trait                                                                                                                         & Disagreeable Personality Trait                                                                                               \\ \specialrule{1.5pt}{0pt}{0pt} 
Speech volume             & 90\% (of the max)                                                                                                                          & 60\% (of the max)                                                                                           \\
Speech speed              & 100\% (of the max)                                                                                                                         & 80\% (of the max)                                                                                           \\
Speech pitch              & 80\% (of the max)                                                                                                                          & 60\% (of the max)                                                                                           \\
Speech style (vocabulary) & \begin{tabular}[c]{@{}c@{}}- Great! I’m moving the first piece of this tower now!\\ - I'll get right to work on this new task!\end{tabular} & \begin{tabular}[c]{@{}c@{}}- I'll move the piece\\ -Alright, if I really have to, I’ll move it.\end{tabular} \\
Feedback given            & \begin{tabular}[c]{@{}c@{}}- Done! It’s great working together!\\ - We're getting close to the finish line!\end{tabular}                    & \begin{tabular}[c]{@{}c@{}}- It's boring moving the pieces\\ - I hope this is over soon\end{tabular}         \\
Gestures                  & Speaking contextual movements                                                                                                               & No gestures                                                                                                  \\
Body movement             & The robot stands up and gestures while speaking.                                                                                            & The robot remains seated throughout the entire interaction.                                                  \\ 
Maintaining eye contact             & Yes                                                                                            & No                                                  \\ 
\specialrule{1.5pt}{0pt}{0pt} 
\end{tabular}
\end{adjustbox}
\end{center}
\end{table*}

The present study considered three main characteristics of the humanoid robot NAO: competence, autonomy, and personality traits. The robot was considered competent when it successfully completed the task; conversely, it was considered not competent when it failed to accomplish the task. Similarly, two levels of autonomy were established, according to the theory of Social Adjustable Autonomy formalized in~\cite{falcone2001human}: one literal and one over-helping. In the literal help mode, the robot performed only the actions explicitly delegated by the user, following a step-by-step process in the construction of the tower. This meant the robot moved each piece only as directed by the user, from the start to the completion of the tower. Conversely, the over-helping mode referred to a level of task adoption that exceeded the user's explicit requests. In this context, the robot not only completed the delegated action (e.g., moving a specific piece) as instructed but also anticipated and performed additional actions that the user did not specifically request (e.g., moving both the delegated piece and the subsequent piece). In fact, the literal level implies a lack of autonomy on the part of the robot, while the over-helping level implies a degree of autonomy.

Finally we designed two types of robot personality traits: agreeable personality and disagreeble personality. Table \ref{tab:PersonalityDesign} summarizes the NAO robot features exploited in our study to implement traits of agreeableness. We based the personality design on previous research about NAO and Pepper robots, as they share multiple technical and behavioral characteristics, being built by the same company.~\cite{chen2024effects, esteban2022should,mileounis2015creating,van2013meet}. The agreeable robot was highly active, speaking with a loud and high-pitched voice, used positive, collaborative, and polite vocabulary, and provided feedback on its experience of collaborating with the human. The robot’s utterances were accompanied by contextual gestures and it maintained eye contact with the human. In contrast, the disagreeable robot spoke with a lower-pitched voice, and a slower speech rate. The phrases used were shorter, and the vocabulary was less collaborative and more hostile. Additionally, the feedback given to the user emphasized its reluctance to collaborate and its perception of performing tasks just because they were instructed. This robot did not gesture while speaking.
Regardless of personality, the robot lowered itself to press the physical button on the console. However, while the agreeable robot stood up each time it engaged with the user, utilizing body language to communicate, the disagreeable robot remained seated throughout the interaction, providing body language cues that were hostile to collaboration with the human. We conducted a preliminary pilot study to evaluate how human users perceived the two polarized personality traits as designed in our experiment. Results confirmed that participants clearly distinguished between them.

\subsection{Participants}\label{subSec:Participants}

A web-based experimental paradigm was administered on Qualtrics.com to a stratified sample of Italian native speakers. 500 participants were recruited, ensuring a balanced representation by gender, age, categories ($<35; 36-55; >55$), geographical regions (Northern, Central, Southern Italy), and educational background (high school or lower, bachelor’s degree or higher). Data collection took place in July 2024 and was performed by the market research firm Bilendi (www.Bilendi.com) that invited and incentivized participants. 127 participants were omitted from all the analyses based on the following exclusion criteria: i) 65 participants were excluded for finishing the survey too quickly (overall completion time $<300$ sec.); ii) 30 participants were removed for responding too rapidly to the likert scale items (RT $<20$ sec.); iii) 33 participants failed the attention check (i.e., a specially designed item that required a specific response). These participants were excluded from all statistical analyses. The final sample included 373 participants ($F = 162; M_{age}: 44.19, SD = 13.14$). Participants gave explicit informed consent after being informed about the study’s objectives and purposes.

\subsection{Measures} \label{sec:Measures}

The present analysis involved the following dependent variables:
\begin{itemize}
    \item Cognitive trust: to assess participants cognitive trust in the robot, we employed a 4-item scale including Competence, Decision-making capability, Task alignment, and Adherence to instructions. 
    \item Affective trust: To assess participants affective trust in the robot, we employed a 4-item scale including Friendliness, Pleasantness, Protectiveness, Synergy.
\end{itemize}
The items used for these measurements were selected from Schaefer's Trust Perception Scale-HRI~\cite{schaefer2013perception}. Schaefer proposed four categories of trust in robots: propensity to trust, affect-based trust, cognition-based trust, and trustworthiness. Consequently, we selected this scale to identify a minimal set of items suitable for measuring both cognitive and affective trust. 
In both scales, participants responded using a 6-point Likert scale, ranging from "very low level of" (1) to "very high level of" (6).
\begin{itemize}
    \item Task Delegation: To assess participants willingness to delegate a new task to the robot based on the concluded interaction, we employed a single-item scale. Given a new task to accomplish (building a new tower), participants were presented with the following question: "Based on the interaction that just concluded, and knowing that the robot understands your goal, to what extent would you guide it in building the tower?" Participants could choose from five incremental levels of control, ranging from completely autonomous (1) to fully controlled (5). For instance, if the goal was to build a tower consisting of three blue cylinders and one red cube, each participant could choose to have the robot autonomously build the tower without explicitly delegating any actions (i.e. Level 1), or to delegate the movement of one piece and then allow the robot to continue building the tower autonomously (i.e. Level 2), or to delegate the movement of two pieces (i.e. Level 3), and so on, up to explicitly delegating the movement of all pieces of the tower, without leaving any decision-making autonomy to the robot (Level 5). It is important to note that less robot autonomy meant more control by the user, that is, lower trust behavior as intentional act of task delegation.
\end{itemize}
Furthermore, we assessed the participants attitude towards robots by exploiting a short version of the 14-items Negative Attitude Towards Robots (NARS) scale~\cite{nomura2006measurement}. Participants responded using a 6-item Likert scale, ranging from strongly disagree (1) to completely agree (6). Each assessment was administered after the interaction with the robot concluded.

\subsection{Statistical notes} \label{subSec:StatNotes}

Statistical analyses were implemented in R~\cite{JSSv067i01}. A Generalized LM model was run for each dependent variable. A significance threshold of $\alpha=.05$ was adopted. \textit{P-values} are reported with the unstandardized estimates ($\beta$), the $95\%$ Confidence Intervals (CIs), and the standard errors (SE). Sociodemographic characteristics (i.e., gender, age, level of education) were included as fixed effects in the models. Subsequently, these variables were removed to enhance model parsimony and model fit in terms of AIC and BIC values, based on non-significant main effects and interactionsz~\cite{JSSv067i01}. Post-hoc comparisons \textit{p-values} were corrected using the Bonferroni correction and were investigated upon a significant main effect or interaction. Their effect sizes are reported using Cohen’s d from emmeans package~\cite{emmeans}. As regard the path analysis, standardized beta coefficients ($\beta$), the relative Standard Error (S.E.), the statistical significance (p-value), and a $95\%$ Confidence interval ($95\%$ CI) were computed for each explanatory variable. The $R^2$ value was calculated for outcome variables~\cite{kline2023principles}. A mediation analysis was conducted, examining the significance of indirect effects through a bootstrapping procedure, including 1000 bootstrapped samples~\cite{byrne2010structural}. Normal distribution was assessed using z scores~\cite{kim2013statistical} and Maximum Likelihood was chosen as the parameter estimation method~\cite{nevitt2001performance}. A sensitivity power analysis was performed~\cite{kumle2021estimating} to ensure that our sample size ($N=373$) was sufficient to detect the effects of interest with adequate statistical power. This analysis proved that our models were sufficient to detect a significant effect as low as $d=25$ with $power=80\%$.
\section{Results}\label{sec:Results}

To evaluate the impact of Robot’s Competence (Non competent / Competent), Autonomy (Non autonomus / Autonomous) and Personality (Disagreeable / Agreeable) on Cognitive and Affective Trust and users Task Delegation, we performed three Generalized Linear Models on Cognitive and Affective Trust and Task Delegation. Attitude towards robot scale was included as covariate in each model. 
\begin{figure}[!t]
\centering
\includegraphics[width=0.6\columnwidth]{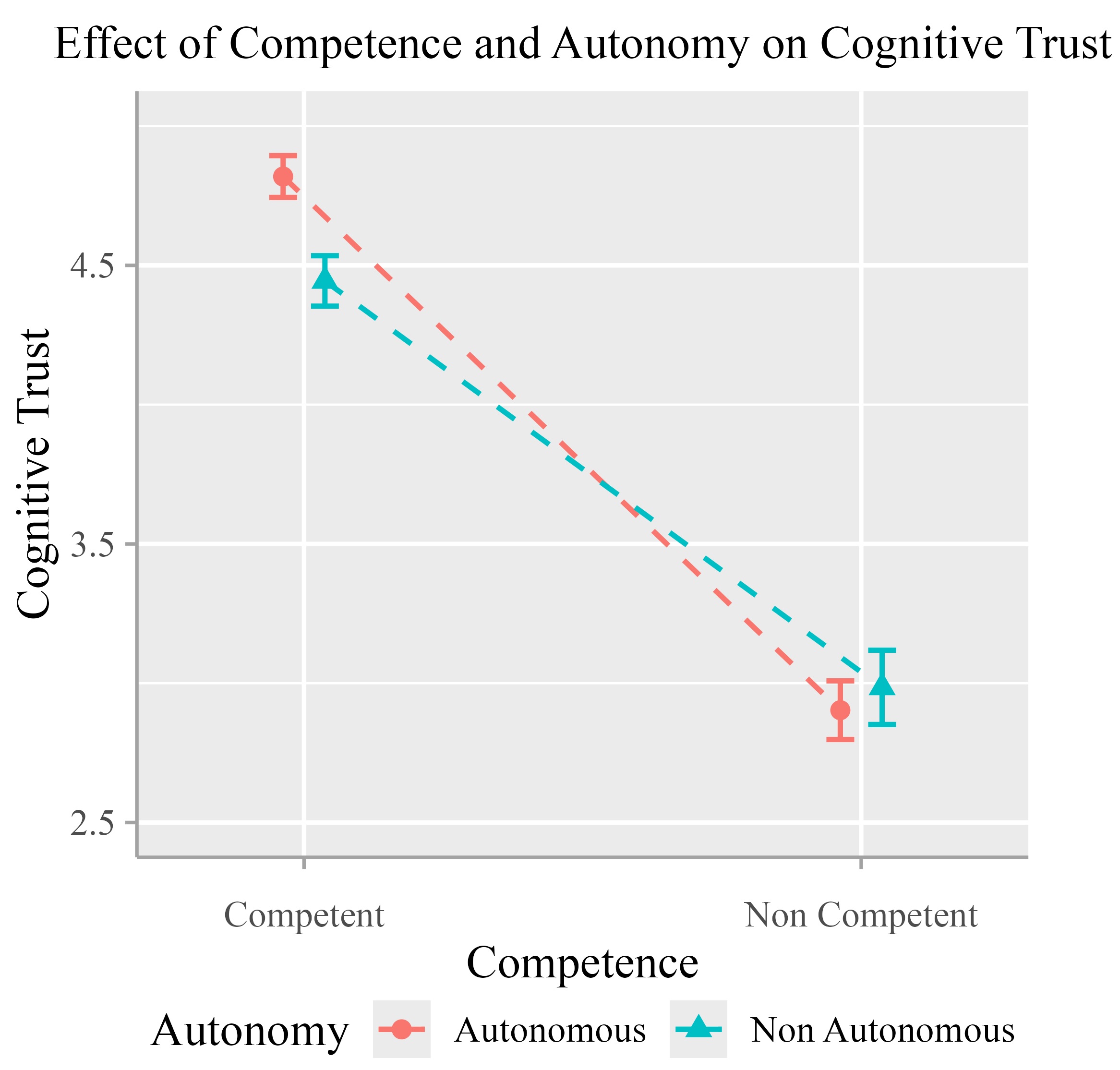}
\caption{Line chart of the interaction effect of robot Autonomy and Competence on Cognitive Trust. When the robot was competent, participants
showed higher cognitive trust when the robot was autonomous.}\label{fig:cog_trust}
\end{figure}
\subsection{Cognitive Trust}

A significant effect of the Competence was found ($\beta=1.66$, $SE=.10$, $p<.001$, $95\%CI [1.46, 1.84]$, $d=1.72$). When the robot completed the task competently, participants reported higher cognitive trust ($M=4.61;SE=.07$) compared to when the robot failed the task ($M=2.96;SE = .07$). The effect of Autonomy ($p=.223$), the effect of Personality ($p=.389$), and their interaction ($p=.276$) were not significant. Conversely, we found a significant Competence × Autonomy interaction ($\mathcal{X}^2(1)=4.24;p=.039$). It suggests that the effect of autonomy on the robot’s cognitive trust varied depending on the robot’s competence level (Figure \ref{fig:cog_trust}). More specifically, when the robot was not competent, the autonomy did not seem to significantly influence cognitive trust ($MLH=3.00$, $SELH=.11$; $MOH=2.92$, $SEOH=.09$; $p=576$). In contrast, when the robot was competent, participants showed higher cognitive trust when the robot was autonomous ($MOH=4.77$, $SEOH=.10$) compared to when it provided no autonomy ($MLH=4.45$, $SELH=.11$; $p<.001$; $d=.29$). This suggests that the robot’s competence is a leading factor in cognitive trust. Lastly, main effect of ATR was significant ($\beta=.33$, $SE=.05$, $p<.001$, $95\%CI [.23,.43]$). 

\subsection{Affective Trust}

\begin{figure}[!t]
\centering
\includegraphics[width=0.6\columnwidth]{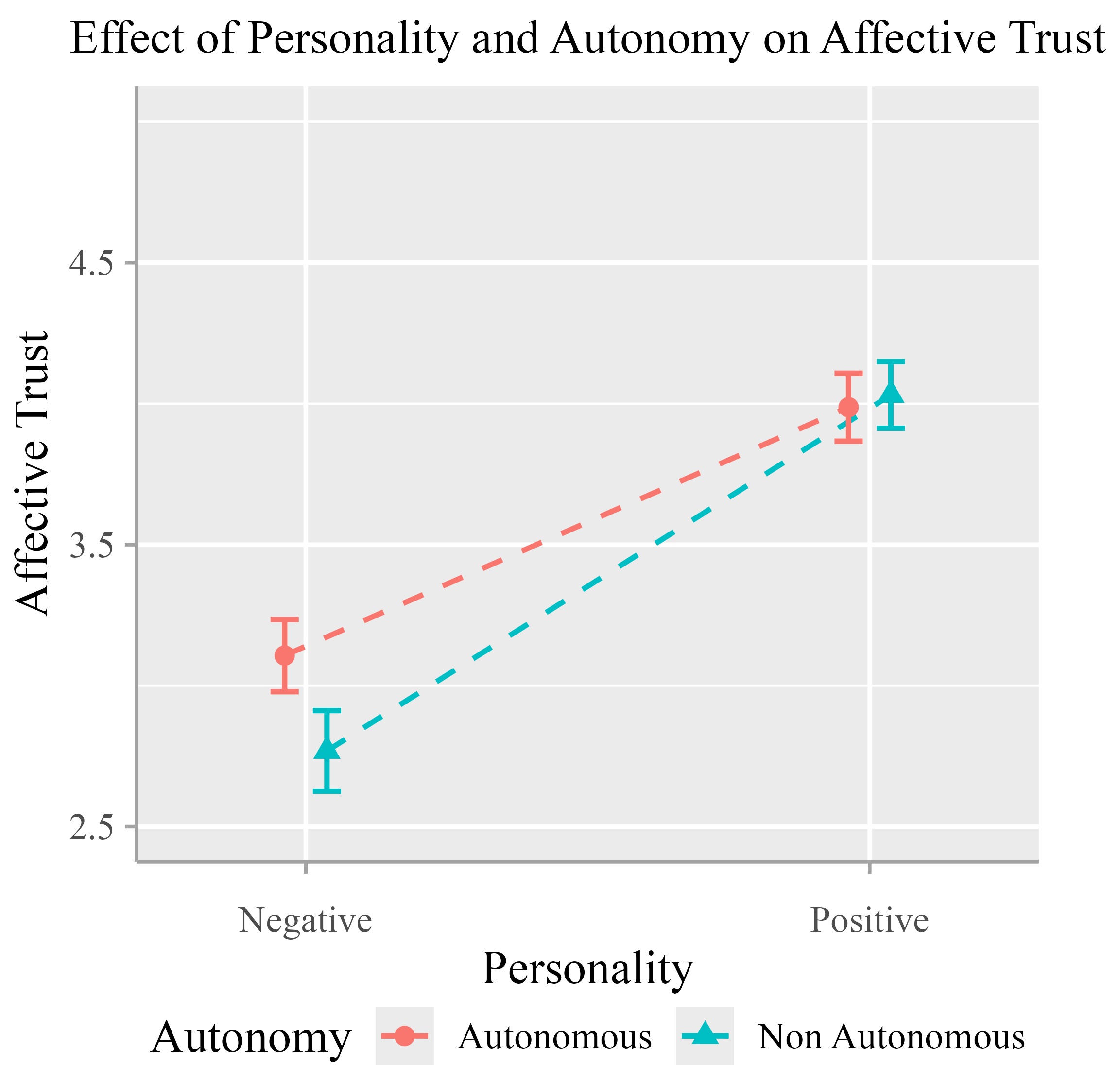}
\caption{Line chart of the interaction effect of robot Autonomy and Personality on Affective Trust. When the robot was disagreeable, autonomy resulted in higher affective trust.}\label{fig:aff_trust}
\end{figure}
A significant effect of Attitude Toward Robots (ATR) was found ($\beta = 0.49$, $SE = .06$, $p <. 001$, $95\% CI [0.39, 0.60]$, $d = .66$), indicating that a more positive attitude towards robots was associated with higher affective trust. Similarly, the effect of Personality was significant ($\beta = 1.06$, $SE = .11$, $p < .001$, $95\% CI [0.85, 1.28]$), with participants reporting greater affective trust for a robot with a pleasant personality ($M = 4.01$, $SE = .08$) compared to a robot with an unpleasant personality ($M = 2.95$, $SE = .08$). Competence also had a significant effect ($\beta = .76$, $SE = .11$, $p<.001$, $95\% CI [.54, .97]$), with higher affective trust reported when the robot was competent ($M = 3.86$, $SE = .08$) compared to when it was non competent ($M = 3.10$, $SE = .08$). The effect of Autonomy was not significant ($p = .272$), and the interactions between Autonomy and Competence ($p = .652$) as well as Personality and Competence ($p = .357$) were also not significant. However, there was a significant Personality × Autonomy interaction ($\mathcal{X}^2 (1) = 4.53$; $p=.033$). 
This interaction indicates that the effect of autonomy on affective trust varied depending on the robot personality (Figure \ref{fig:aff_trust}). Specifically, when the robot was disagreeable, autonomy resulted in higher affective trust ($MLH = 3.13$, $SELH = .10$) compared to when the robot was non autonomous ($MOH = 2.77$, $SEOH = .12$; $p = .024$, $d = .32$). Conversely, when the robot had a agreeable personality, the difference between autonomy and non autonomy was not significant ($MLH = 4.07$, $SELH = .11$; $MOH = 3.96$, $SEOH = .11$; $p = .024$; $p = .464$). In summary, the robot’s personality moderated the effect of autonomy on affective trust. A disagreeable robot benefited more from autonomy in terms of affective trust, while an agreeable robot did not show significant differences between the types of help. This suggests that for robots with disagreeable personality, the perception of autonomy may mitigate the negative impact of their personality on affective trust, whereas for robots with agreeable personality, the type of help did not play a crucial role. 

\subsection{Task Delegation}

\begin{figure}[!t]
\centering
\includegraphics[width=0.8\columnwidth]{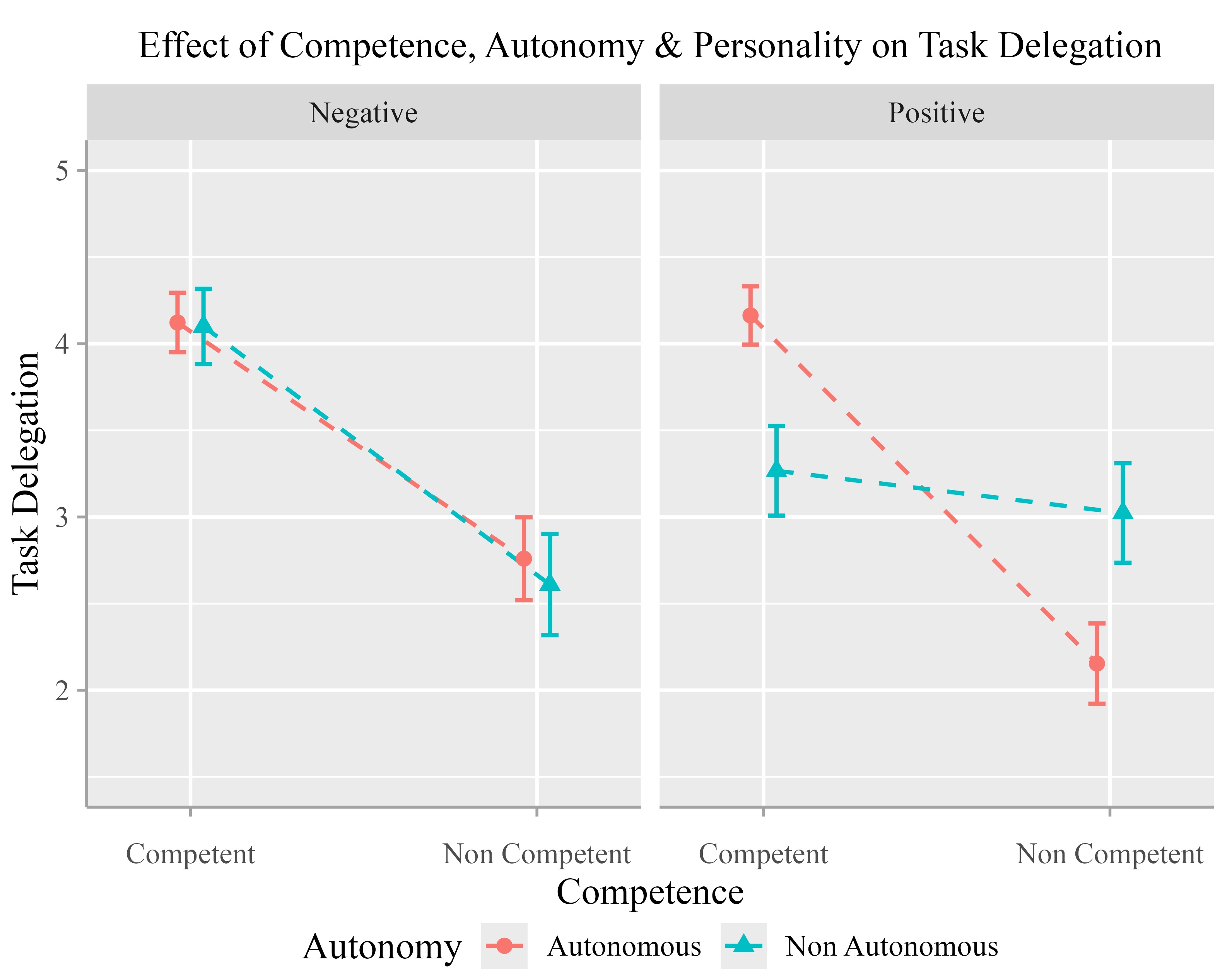}
\caption{Line chart of the interaction effect of robot Competence, Autonomy and Personality on Task Delegation. Participants valued a competent and autonomous robot only when it was agreeable, while disagreeableness made them indifferent to its autonomy or competence.}\label{fig:task_del}
\end{figure}
\begin{figure}[!t]
\centering
\includegraphics[width=0.8\columnwidth]{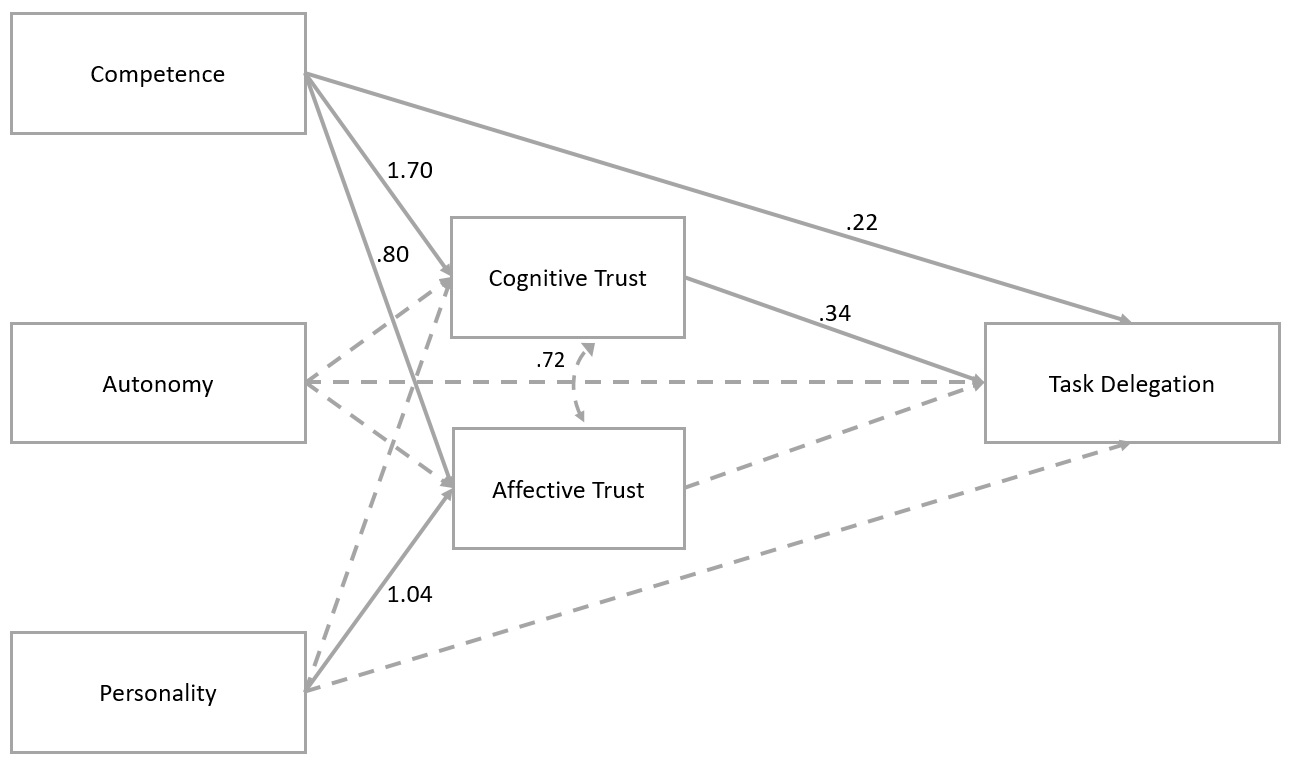}
\caption{Path analysis results. Parameters estimates are standardised. Dotted lines represent insignificant paths. Continuous lines represent significant paths.}\label{fig:resultsPath}
\end{figure}
A significant effect of Competence on Task Delegation was found ($\beta=1.27$, $SE=.17$, $p<.001$, $95\% CI [.94, 1.60]$, $d=.88$), indicating that participants required more guidance when the robot was non competent ($M=2.64$,$SE=.12$) compared to when the robot was competent ($M=3.91$, $SE=.12$). Neither Attitude Toward Robots ($p=.449$), Personality ($p=.141$), nor Autonomy ($p=.794$) had significant main effects. However, a significant two-way interaction was found between Autonomy and Competence ($\mathcal{X}^2(1)=5.84$, $p=.016$). Participants showed higher levels of task delegation when the robot was competent and autonomous ($M=4.13$, $SE=.16$) compared to when it was competent but non autonomous ($M=3.69$, $SE=.17$). Contrarily, when the robot was non competent, participants showed higher levels of delegation when the robot was non autonomous ($M=2.82$, $SE=.18$) as compared to when it acted autonomously ($M=2.46$, $SE=.16$). Despite the non-significance of these post-hoc comparisons, the overall significant interaction suggests that autonomy’s influence on task delegation was moderated by the robot’s competence. 
Additionally, a significant three-way interaction between Personality, Autonomy, and Competence was observed ($\mathcal{X}^2(1)=8.08$, $p=.004$). This interaction reveals that the impact of autonomy on task delegation varied depending on both the robot’s Competence and Personality (Figure \ref{fig:task_del}). Specifically, for the disagreeable robot, there were no differences in task delegation when the robot was competent or non competent between autonomous ($Mcom=4.12$, $SEcom=.23$; $MIncom=2.76$, $SEIncom=.22$) and non-autonomous robots ($Mcom = 4.10$, $SEcom = .25$; $MIncom = 2.25$, $SEIncom = .23$; $ps > .630$). Conversely, for the agreeable robot, participants showed greater task delegation when the robot was autonomous and competent ($M = 4.15$, $SE = .23$) compared to when it was non-autonomous and competent ($M = 3.27$, $SE = .24$; $p = .008$). Symmetrically, a non-competent agreeable robot received a higher task delegation when in was non autonomus ($M = 3.03$, $SE = .25$) as compared to when it exhibited an autonomous behavior ($M = 2.16$, $SE = .22$; $p = .008$). In short, when the robot had a disagreeable personality, participants were not influenced by either autonomy or competence when deciding how much to delegate to the robot. Conversely, when the robot had a agreeable personality, autonomy was more appreciated if the robot was competent, but the lack of competence made a more structured and predictable guidance preferable.

\subsection{Path Analysis: Examining the Predictive and Mediating Roles of Affective and Cognitive Trust} \label{subSec:PathAnalysis}

To better understand the predictive roles of affective and cognitive trust in determining task delegation to the robot, and to examine their mediating effects between the predictors (Personality, Autonomy, Competence) and task delegation, we conducted a path analysis (Figure \ref{fig:resultsPath}). This analytical approach allowed us to simultaneously assess the direct relationships between the predictors and task delegation, the indirect relationships mediated by affective and cognitive trust, as well as to quantify the relative importance of affective and cognitive trust in the decision-making process regarding delegation. The results showed that the hypothesized predictors collectively explained 18.3\% of the variance in task delegation ($\mathcal{R}^2=.183$; $95\% CI [.116, .258]$). Additionally, the model explained 42.9\% of the variance in cognitive trust ($\mathcal{R}^2 = .429$; $95\% CI [.352, .503]$) and 25.0\% of the variance in affective trust ($\mathcal{R}^2 = .250$; $95\% CI [.177, .328]$). As regards the direct effects, Personality had a significant positive effect on affective trust ($\beta = 1.043$, $SE = .120$, $p < .001$; $95\% CI [.807, 1.279]$), but it did not significantly predict cognitive trust ($\beta = .075$, $SE = .103$, $p = .466$). Autonomy did not significantly affect either cognitive ($\beta = -.145$, $SE = .104$, $p = .163$) or affective trust ($\beta = -.162$, $SE = .121$, $p = .179$). Competence significantly affected both cognitive trust ($\beta = 1.708$, $SE = .102$, $p < .001$;$ 95\% CI [1.508, 1.907]$) and affective trust ($\beta = .819$, $SE = .120$, $p < .001$; $95\% CI [.584, 1.054]$).  Moreover, the results showed that Competence also significantly predicted task delegation ($\beta = .704$, $SE = .228$, $p = .002$; $95\% CI [.257, 1.150]$), indicating that higher competence was associated with higher levels of delegation. However, personality ($\beta = -.296$, $SE = .191$, $p = .120$), and autonomy ($\beta = .002$, $SE = .167$, $p = .990$) did not have significant direct effects on delegation. Finally, only cognitive trust had a significant positive effect on delegation ($\beta = .345$, $SE = .105$, $p < .001$; $95\% CI [.140, .550]$), suggesting that higher cognitive trust in the robot led to higher levels of delegation. On the contrary, affective trust was not associated with task delegation ($\beta = .011$, $SE = .087$, $p = .902$).  Indirect effect analyses revealed that competence indirectly influenced delegation through cognitive trust ($\beta = .589$, $SE = .179$, $p < .001$; $95\% CI [0.239, 0.940]$), indicating a significant mediation effect. Specifically, higher competence was associated with higher cognitive trust ($\beta = 1.708$, $SE = 0.102$, $p < .001$), which in turn led to higher delegation. All other indirect effects were not significant.  This analysis highlighted that while personality and autonomy played roles in shaping trust, it was the actual competence of the robot that most significantly drove both cognitive trust and the ultimate decision to delegate tasks. Cognitive trust, rather than affective trust, was the key mediator in this process. 

\section{Discussion}\label{sec:Discussion}
The goal of this work was to expand the existing literature on Human-Robot Trust, by understanding how different levels of robot attributes influence human trust attitude and behavior. Next, we briefly discuss each of these aspects along with some possible implications.

First of all, the robot level of competence in achieving a delegated task influenced the level of cognitive trust in the participants who delegate the task (\textbf{H1}). In particular, when the robot successfully completed the task, participants cognitive trust was significantly higher compared to when the robot fails. The effect size (d = 1.72) showed the main role of competence in building cognitive trust. This was in line with the literature on trust in HRI, which consider robot competence (e.g. task performance) to be the most important determinant for building trust attitude~\cite{xie2019robot,hancock2021evolving}, in particular cognitive trust~\cite{gompei2018factors,chen2024effects,lim2022we}. Furthermore, finding concerning the relationship between personality traits and cognitive trust confirmed hypothesis \textbf{H2}, aligning our study to the literature~\cite{hancock2011meta,gompei2018factors}.

A more interesting finding was that, unlike robot competence, the robot autonomy and personality do not significantly affect cognitive trust. More specifically, this indicated that autonomy alone did not determine participants cognitive trust~\cite{hancock2021evolving,gompei2018factors}. While the results seemed to confirm that autonomy alone did not have an effect, they showed a interaction between autonomy and competence. In particular, in line with hypothesis \textbf{H3}, robot autonomy moderated robot competence. When the robot was not competent, autonomy did not significantly influenced cognitive trust, as participants showed similar levels of trust regardless of whether the robot acted exactly as delegated (literal-help mode) or performed actions without the participant's explicit delegation (over-help mode). Vice versa, participants reported higher cognitive trust when the competent robot acted autonomously, compared to when it followed their instructions step-by-step. This result suggested that the interaction with competence is a prerequisite for autonomy to positively influence cognitive trust: participants were more likely to trust a competent robot if it demonstrated autonomy in task achievement. This is an important finding, consistent with our hypothesis (\textbf{H3}) that autonomy impacts cognitive trust only in relation to the robot competence and the expectations associated with the delegated task, rather than as an absolute dimension, as previous studies suggested~\cite{gompei2018factors,stapels2022robocalypse,cantucci2023cognitive}.

Affective trust toward the robot was positively influenced by robot personality, rather that robot autonomy considered alone. Personality traits played a fundamental role in the affective trust building process~\cite{chen2024effects,lim2022we}. Our findings confirmed that participants showed greater affective trust when the robot exposed high levels of agreableness traits. Competence also significantly affected affective trust, with higher trust reported for competent robots compared to non competent ones. This finding supported the notion that perceived ability is crucial for affective trust formation in HRI, as well as for cognitive trust (\textbf{H2}). 

Similarly to cognitive trust, the results also showed that the robot autonomy alone did not directly influenced affective trust, contrary to what was suggested by previous studies~\cite{gompei2018factors}, which indicated that autonomy itself is a determinant of affective trust towards robots. However, our study provided more in-depth evidence, revealing a significant interaction between the robot personality and autonomy, which showed that the impact of autonomy on affective trust was in relation with the robot personality. For the disagreeable robot, autonomy led to higher affective trust. Conversely, for the agreeable robot, the difference between autonomy and non autonomy was not significant. This interaction suggested that autonomy can mitigate the negative impact of a personality trait on affective trust. In this case, being autonomous may present a compensatory mechanism, potentially by demonstrating willingness to interact with participants, or reducing perceived negative traits. In contrast, for agreeable robots, the type of help did not significantly alter affective trust, indicating that a positive personality may be sufficient to maintain affective trust irrespective of autonomy levels. This result is important, and extend the role of autonomy as mediator between affective trust and robot personality. Similarly to the cognitive trust findings, autonomy per se does not enhance emotion- and cognition-based trust; however, when combined with other attributes like competence and personality, it amplifies their positive impact.

Task delegation is considered the action resulting by the human intention to trust a robot~\cite{castelfranchi2010trust}. It is considered a human behavior, linked to the expactations on robot characteristics like competence, autonomy, and personality traits ~\cite{soh2020multi,schaefer2013perception}. Behavioral manifestations of trust during human robot interaction, based on trust attitudes, have received less experimental focus than trust attitude analysis~\cite{xie2019robot,roesler2024dynamics,soh2020multi}. The present study tried to extend literature in human-robot trust analyzing the effects of robot competence, autonomy, and personality on participants trust behavior, corresponding to task delegation.

As expected, the results indicated a significant main effect of robot competence on task delegation (\textbf{H1}). The main effect found ($d = 0.88$) demonstrated the influence of the robot competence in achieving an appropriate balance between trust and control: participants exercised explicit control over the robot actions when it proved non competence in task achievement, compared to when it was competent, delegating greater autonomy. Furthermore, the findings confirmed the predominant role of competence in task achievement on overall trust~\cite{hancock2011meta,roesler2024dynamics,castelfranchi2010trust}, encompassing both cognition- and affect-based evaluations, as well as human behavior~\cite{xie2019robot,soh2020multi}.

Similar to the cases of cognitive and affective trust, no significant main effects of personality or autonomy per se were found on task delegation. The impact of these two attributes, when considered individually, was confirmed to be non-significant. However, we observe that the interaction effects between the robot attributes revealed a more interesting and complex picture.
First of all, a significant interaction between autonomy and competence was observed. Participants showed higher levels of task delegation when the robot was both competent and autonomous, compared to when it was competent but less autonomous. Conversely, when the robot was non competent, higher task delegation was observed when the robot acted as instructed rather than acting autonomously. This interaction highlighted that autonomy's impact on task delegation depends on the robot competence, suggesting a moderation effect (\textbf{H3}): autonomy was valued when paired with competence, but may be less effective or even counterproductive associated with a lower competence~\cite{capiola2023you,chiou2023trusting,stapels2022robocalypse}.

Furthermore, a significant and interesting interaction was found between all the robot attributes considered in this work. More specifically, findings suggested that negative personality traits may negatively influence the effects of competence and autonomy, potentially due to a perceived lack of robot willingness to collaborate. In contrast, when the robot exhibited agreeable traits, competence and autonomy significantly influenced task delegation. In particular, autonomy again played a moderating role between competence and task delegation. These findings suggest that robot personality significantly influences how autonomy and competence affect task delegation. When the robot was agreeable, participants delegated more when the robot exhibited both competence and autonomy. Conversely, when they interacted with a disagreeable robot, participants were more prescriptive by default. 

The path analysis described in section \ref{subSec:PathAnalysis} confirmed and suggested new findings.
Results confirmed robot competence the main determinant for trust dimensions investigated in this work. Competence had a significant positive effect on both cognitive and affective trust, and more importantly, it predicted task delegation decision. The relationship between competence and task delegation was mediated by cognitive trust, reinforcing the notion that competence enhances cognitive trust, which in turn fosters task delegation. The indirect effect analysis supported this, revealing a significant mediation effect of cognitive trust in the relationship between competence and delegation.

In contrast, personality and autonomy did not demonstrate significant direct effects on task delegation. While personality positively impacted affective trust, it did not significantly influenced cognitive trust or task delegation. Similarly, autonomy had no effect on either trust dimension or delegation. These results suggested that while personality and autonomy may shape trust, their impact on delegation is relatively limited compared to competence. This finding aligns with the notion that the functional attributes of the robot, such as competence, play a more fundamental role in influencing delegation decisions than the broader, more abstract constructs of personality and autonomy.

Cognitive trust, rather than affective trust, emerged as the key mediator in the delegation process. Our results showed that only cognitive trust had a significant positive effect on delegation, whereas affective trust did not. This suggests that users' judgments regarding the robot capability, captured by cognitive trust, are more decisive in their decision to delegate tasks than their emotional feelings of trust towards the robot. This distinction highlights the practical significance of cognitive trust in facilitating effective human-robot interaction.

\subsection{Limitations}

The study was conducted online, which reduced uncertainties related to physical spaces and enabled broad accessibility to a diverse and bigger participant sample. However, the virtual interaction approach may not have fully captured the complexity of in-person interactions. In a physical setting, participants might have exhibited different behaviors, potentially leading to divergent findings. Nonetheless, in HRI research, the traditional approach of presenting physical robots in physical settings is increasingly complemented by the use of virtual representations (e.g., videos, virtual reality, etc.)~\cite{Esterwood2024}. Previous studies~\cite{you2017emotional,plomin2023virtual,rossi2023evaluating} have explored this approach as a valid alternative for assessing users' trust and engagement with robots.

This study primarily relied on subjective trust assessments (i.e., questionnaires), a methodology widely established in this type of research. However, incorporating behavioral measures would offer a more comprehensive understanding of how humans trust is reflected in their behavior. Future research could integrate objective behavioral and physiological measures~\cite{alzahrani2024real,campagna2025systematic} to complement self-reported data and enhance the robustness of trust assessments.

\section{Conclusions}\label{sec:Conclusions}

In this paper we examined the influence of robot competence, autonomy and personality traits on different dimensions of people trust, namely cognitive trust, affective trust and behavioral trust (i.e. task delegation). We presented an online experiment in which 373 participants were asked to interact with the humanoid robot Nao for accomplishing a construction task. The interaction was structured through a sequence of task delegation and task adoption. We found that robot competence is a crucial factor for predicting cognitive, affective, and task delegation. Conversely, while personality traits of robots were found to significantly influence affective trust, they did not affect cognitive trust or task delegation, indicating that emotional perceptions toward robots may be less consequential for practical task-related decisions. Furthermore, the moderating role of autonomy highlights its importance in enhancing the impact of competence on cognitive trust and personality on affective trust. This implies the need for careful design and calibration of autonomy levels in robots to enhance human trust. Results also show that cognitive trust affects task delegation, suggesting that humans are more likely to assign tasks to robots they perceive as competent, whereas affective trust did not significantly influence this behavior. These insights have important implications for the design of robotic systems, particularly in environments where trust is essential for effective human-robot collaboration.

\bibliographystyle{unsrt}
\bibliography{references}

\end{document}